\documentclass[
 reprint,
 amsmath,amssymb,
 aip,
]{revtex4-2}

\usepackage{graphicx}
\usepackage{subfig}
\usepackage{dcolumn}
\usepackage{bm}
\usepackage{caption}
\usepackage{hyperref}

\begin{document}

\title{Recovering Non-Maxwellian Velocity Distribution Functions from Collective Thomson Scattering Spectra Using Generalized Basis Representations}

\author{Greg A. Baker}
\thanks{These authors contributed equally to this work and are listed alphabetically.}
\author{Bradyn Klein}
\thanks{These authors contributed equally to this work and are listed alphabetically.}
\author{Kristopher Wright}
\affiliation{Department of Physics and Astronomy, University of California, Los Angeles, CA 90095, USA}

\author{Vedang Bhelande}
\affiliation{Department of Physics and Astronomy, University of Rochester, Rochester, NY 14627, USA}

\author{Zhaoyan Sun}
\author{Rayner Sutardja}
\affiliation{Department of Physics and Astronomy, University of California, Los Angeles, CA 90095, USA}

\author{Samuel R. Totorica}
\affiliation{Department of Astrophysical Sciences, Princeton University, Princeton, NJ 08544, USA}
\affiliation{Department of Astro-fusion Plasma Physics (AFP), Headquarters for Co-Creation Strategy, National Institutes of Natural Sciences, Tokyo 105-0001, Japan}

\author{Derek B. Schaeffer}
\email{dschaeffer@physics.ucla.edu}
\affiliation{Department of Physics and Astronomy, University of California, Los Angeles, CA 90095, USA}

\begin{abstract}
\begin{center}
\textbf{ABSTRACT}
\end{center}
Thomson scattering (TS) is a key diagnostic for inferring plasma parameters from scattered power spectra, but its interpretation depends strongly on assumptions about the underlying velocity distribution functions (VDFs). Conventional analyses typically assume Maxwellian distributions, which can be invalid in high-energy-density and laboratory astrophysics plasmas where collisionless, multi-population dynamics produce non-thermal features. In such cases, Maxwellian models may reproduce observed spectra while yielding inaccurate or misleading plasma parameters. We present a generalized TS analysis framework that represents species-specific VDFs in flexible, physically motivated bases and so reconstructs complex, non-equilibrium distributions. To explore the resulting high-dimensional parameter space, we build the forward model from differentiable operations and optimize the VDF parameters by gradient descent. We apply the method to synthetic TS spectra generated from particle-in-cell simulations of a collisionless shock with interacting hydrogen and carbon populations. Immediately upstream of the shock, where the plasma is most non-thermal, the flexible bases cut the mean normalized ion VDF residual from 38.7\% for a Maxwellian model to 20.3\% and 20.5\%, with correspondingly improved time-resolved plasma parameters. We also introduce a systematic method of uncertainty quantification that reveals degeneracies between ion species, highlighting the importance of both flexible VDF modeling and a statistical mapping of the solution space for this inherently ill-posed inversion problem.
\end{abstract}

\maketitle

\section{Introduction}
\label{sec:introduction}

Thomson scattering (TS) is a powerful diagnostic for characterizing plasmas. Comparing measured spectra to a theoretical model yields fundamental plasma parameters such as electron and ion densities ($n_e, n_i$), temperatures ($T_e, T_i$) and bulk flow velocities ($u_e, u_i$)~\cite{froula2011,glenzer1999}. Accurate time-resolved analysis can provide critical insight into processes such as energy partitioning and flow interactions. The interpretation of TS data, however, depends fundamentally on the underlying velocity distribution functions (VDFs).

Most conventional methods of TS analysis assume Maxwellian distributions, corresponding to thermodynamic equilibrium. However, many plasmas of interest in high-energy-density (HED) science and laboratory astrophysics deviate from this assumption. These systems often exhibit weaker collisionality and consist of multiple interacting populations with distinct velocities and temperatures. Consequently, the true VDF can exhibit multi-peaked structure, asymmetry or non-thermal broadening that no Maxwellian can capture---as in-situ measurements at Earth's bow shock have long shown~\cite{feldman1983,sckopke1990}.

Electron plasma wave (EPW) and ion acoustic wave (IAW) spectral features generated from multi-component plasmas can often be reasonably fit by a Maxwellian model, despite arising from fundamentally different VDFs~\cite{chapman2011,milder2019,saito2000}. However, imposing a Maxwellian model in these regimes can lead to biased or inaccurate inference of plasma parameters, obscuring important dynamics. In such cases, agreement between a Maxwellian fit and the spectra does not guarantee an accurate characterization of the plasma, emphasizing the need for more flexible approaches to TS analysis.

Several approaches already address deviations from Maxwellian behavior in plasma diagnostics: models with multiple Maxwellian components, alternative VDF bases~\cite{zheng1997,saito2000,foo2023}, and machine-learning inference~\cite{pokornik2024,eisenbach2024,vanhoomissen2026}. Each trades off computational cost, robustness on experimental data, and the restrictiveness of the modeling constraints it imposes. Foo \textit{et al.}~\cite{foo2023}, for example, pair an arbitrary forward model with differential evolution (DE)~\cite{storn1997} and Markov Chain Monte Carlo (MCMC)~\cite{foremanmackey2013} fitting, but require the user to supply a parameterized non-Maxwellian VDF in advance. The gap this leaves is concrete: recent magnetized-shock experiments have deferred analysis of the IAW spectrum due to the appearance of non-Maxwellian features~\cite{schaeffer2019shock,valenzuela2026}.

In this work, we introduce a framework for representing species-specific particle VDFs using two distinct basis sets. This technique reconstructs complex, multi-population distributions while remaining both physically motivated and computationally feasible. Applied to TS analysis, it provides a more general mapping between the plasma state and the observed spectra, resolving non-thermal features that conventional modeling assumptions obscure.

To infer the resulting high-dimensional parameter space efficiently, we employ automatic differentiation~\cite{baydin2018} within the forward model, optimizing the VDF parameters by gradient descent. We apply this approach to synthetic Thomson scattering datasets generated from experimentally motivated particle-in-cell (PIC) simulations~\cite{germaschewski2016,fox2018} of a collisionless shock in which a laser-driven hydrogen/carbon piston interacts with a colder hydrogen background.

Compared to the Maxwellian model, both bases recover the underlying hydrogen velocity distributions more accurately at times preceding the shock---roughly halving the mean normalized VDF residual---and so reconstruct the hydrogen flow velocity and temperature more accurately. In regions with fewer non-thermal features they perform comparably to the Maxwellian model. Non-Maxwellian basis sets combined with an arbitrary forward model therefore capture complex, evolving plasma dynamics in non-thermal regions that standard modeling assumptions under-resolve.

We also outline a Monte Carlo method of uncertainty quantification that refits many perturbed initializations to map the solution space~\cite{knapp2023,milder2021stat}. It provides uncertainties on the reconstructed VDFs and plasma parameters and exposes multi-modal solution structure in complex plasma regimes. In this case, the Monte Carlo revealed a degeneracy between hydrogen and carbon inherent to the forward model inversion of a multi-ion plasma, providing insight into the potential limitations of TS analysis for plasmas with multiple ion species.

Together, improved reconstruction of non-thermal features and a statistical map of the solution space are what make this analysis viable for experimental data, where the true plasma state is not known in advance.

The paper is organized as follows. Section~\ref{sec:Methodology} describes the Thomson scattering forward model, the PIC simulation that supplies the synthetic data, the automatic-differentiation optimization scheme, the two generalized basis representations, and the fitting procedure. Section~\ref{sec:spectral_reconstruction} assesses how well each basis reconstructs the spectra, Section~\ref{sec:vdf_reconstruction} extends that assessment to phase space, and Section~\ref{sec:time_resolved_parameters} compares the time-resolved plasma moments. Section~\ref{sec:uncertainty} presents the Monte Carlo uncertainty analysis and the hydrogen/carbon degeneracy it reveals, and Section~\ref{sec:conclusion} summarizes the results and outlines future work.

\section{Methodology}
\label{sec:Methodology}
\subsection{Thomson Scattering Diagnostic Model}
Optical Thomson scattering is a non-perturbative diagnostic in which a high-intensity probe beam is focused onto the plasma region of interest, scattering from charged particles~\cite{froula2011,glenzer1999}. Spectrometers collect the scattered radiation to produce an intensity profile, and fitting its spectral features with a numerical model yields the fundamental parameters of the system: electron density, drift velocities, temperatures, ion fractions and charge states.

The spectra measured in Thomson scattering experiments are described by the spectral density function $S(\mathbf{k}, \omega)$, which characterizes the distribution of scattered power as a function of wavevector $\mathbf{k}$ and frequency shift $\omega$. Here, the scattering wavevector is defined by the probe and scattered light geometry as $\mathbf{k} = \mathbf{k}_s - \mathbf{k}_0 $, where $\mathbf{k}_0$ and $\mathbf{k}_s$ are the incident and scattered photon wavevectors, respectively. The frequency shift $\omega = \omega_s - \omega_0$ corresponds to the difference between scattered and incident photon frequencies and is directly related to plasma motion along $\mathbf{k}$.

The spectral density function for a multi-ion plasma can be written as

\begin{align}
S(\mathbf{k}, \omega) &= \frac{2\pi}{k} \left| 1 - \frac{\chi_e}{\epsilon} \right|^2 f_{e0}\left( \frac{\omega}{k} \right) \nonumber \\
&+ \sum_j \frac{2\pi}{k} \frac{Z_j^2 N_j}{N} \left| \frac{\chi_e}{\epsilon} \right|^2 f_{i0,j}\left( \frac{\omega}{k} \right),
\end{align}

\noindent where $Z_j$ and $N_j$ are the average charge states and number per unit volume of the $j^{th}$ ion species, respectively, such that $N = \sum_j N_j Z_j$. The electron and per-species ion susceptibilities, $\chi_e$ and $\chi_j$, form the longitudinal plasma dielectric function, $\epsilon = 1 + \chi_e + \sum_j \chi_{j}$. The functions $f_{e0}$ and $f_{i0}$ represent the electron and ion VDFs evaluated at the phase velocity $\omega/k$. The first term corresponds to electron scattering contributions, while the second term describes the ion scattering features~\cite{froula2011,hutchinson2002}.

The nature of the observed scattering spectrum is governed by the scattering parameter
\begin{equation}
\alpha = \frac{1}{k \lambda_D},
\end{equation}
\noindent where $\lambda_D$ is the electron Debye length. This parameter determines whether the scattering process probes individual particle motion or collective behavior. In the non-collective regime ($\alpha \ll 1$), the scattering spectrum is dominated by individual electron motion, and the measured signal maps directly onto the electron VDF. By contrast, in the collective regime ($\alpha \gtrsim 1$)~\cite{salpeter1960}, coherent modes become important, and the spectrum exhibits distinct electron plasma wave (EPW) and ion acoustic wave (IAW) features that depend on both electron and ion dynamics. In this work, we focus exclusively on the collective scattering regime, as $1.4 \lesssim \alpha \lesssim 5.3$ for the entire time interval.

\subsection{PIC Simulation of Collisionless Shock}

We test the fitting models developed here on data from 1D particle-in-cell (PIC) simulations of experimentally-relevant laser-driven collisionless shocks on the OMEGA laser facility \cite{schaeffer2019shock,valenzuela2026}.  We chose this system because it contains both strongly non-Maxwellian VDFs and multi-ion VDFs that are challenging to fit with conventional routines.  The simulations model the expansion of a laser-ablated piston plasma into a background plasma embedded in a uniform magnetic field oriented perpendicular to the plasma expansion, similar to those described in \onlinecite{fox2018,schaeffer2020piston,pongkitiwanichakul2024}. The background plasma was composed of hydrogen (H).  The piston plasma was composed of equal parts fully ionized carbon (C$^{+6}$) and hydrogen. Consequently, the plasma at a given location could be composed of up to three different ion populations: a colder background H species, a hotter H species (either from the shock or piston), and a piston C species. Particle VDFs are extracted directly from the PIC simulation and used as inputs to the TS forward model to generate synthetic spectra \cite{foo2023}.

The simulations were performed with PSC~\cite{germaschewski2016}, a fully electromagnetic, relativistic particle-in-cell code that includes a kinetic model of laser ablation~\cite{fox2018}. An inverse bremmstrahlung heating operator provides realistic laser absorption based on local plasma conditions \cite{hyder2026}, resulting in ablation of the piston plasma from a target that self-consistently expands into the pre-initialized, uniform, magnetized ambient plasma.  This reproduces a laser experiment more closely than the counter-streaming or expanding-slab initializations used in classical shock simulations. The runs are quasi-1D in that they contain a few cells with uniform driving conditions in the transverse direction, and use a strictly perpendicular geometry ($\theta_{B0}=90^\circ$) between the piston expansion and the background magnetic field. Following Schaeffer \textit{et al.}~\cite{schaeffer2020piston}, the electron parameters are reduced---an ion-to-electron mass ratio of 100 and a reduced speed of light $c_{sim}/c_{phys}=1/50$---while the ion-scale parameters are kept matched to experiment. Coulomb collisions are included through a Takizuka--Abe binary collision operator, with the electron and ion collisionality matched to typical HED laser experiments \cite{totorica2026}. The upstream electron density was $5\times10^{18}$ cm$^{-3}$, the upstream magnetic field was $10.5$ T, and the upstream electron and ion temperatures were $10$ eV.

All times in this work are reported in units of the inverse upstream ion gyrofrequency $\omega_{ci}^{-1}=1.0\,\mathrm{ns}$.  The simulation data was recorded with 150 discrete temporal snapshots to more closely resemble the resolution of experimental TS data collection, with each time step of the simulation representing $0.077 \,\omega_{ci}^{-1}$; all times are thus integer multiples of this value.

The synthetic spectra are generated using a probe wavelength of 527.0 nm with a scattering angle of $90^\circ$. Collective TS is sensitive only to the projection of the VDF onto $\mathbf{k}$; we orient $\mathbf{k}$ along the shock propagation direction $\hat{z}$, so the synthetic spectra are computed from the reduced distributions $f_s(v_z)$. 10\% Gaussian noise relative to each spectral lineout's maximum is added before fitting to better approximate noisy experimental data. All fits presented in this work are performed on a different noise realization per time step.

\begin{figure}[t]
    \centering
    \includegraphics[width=\columnwidth]{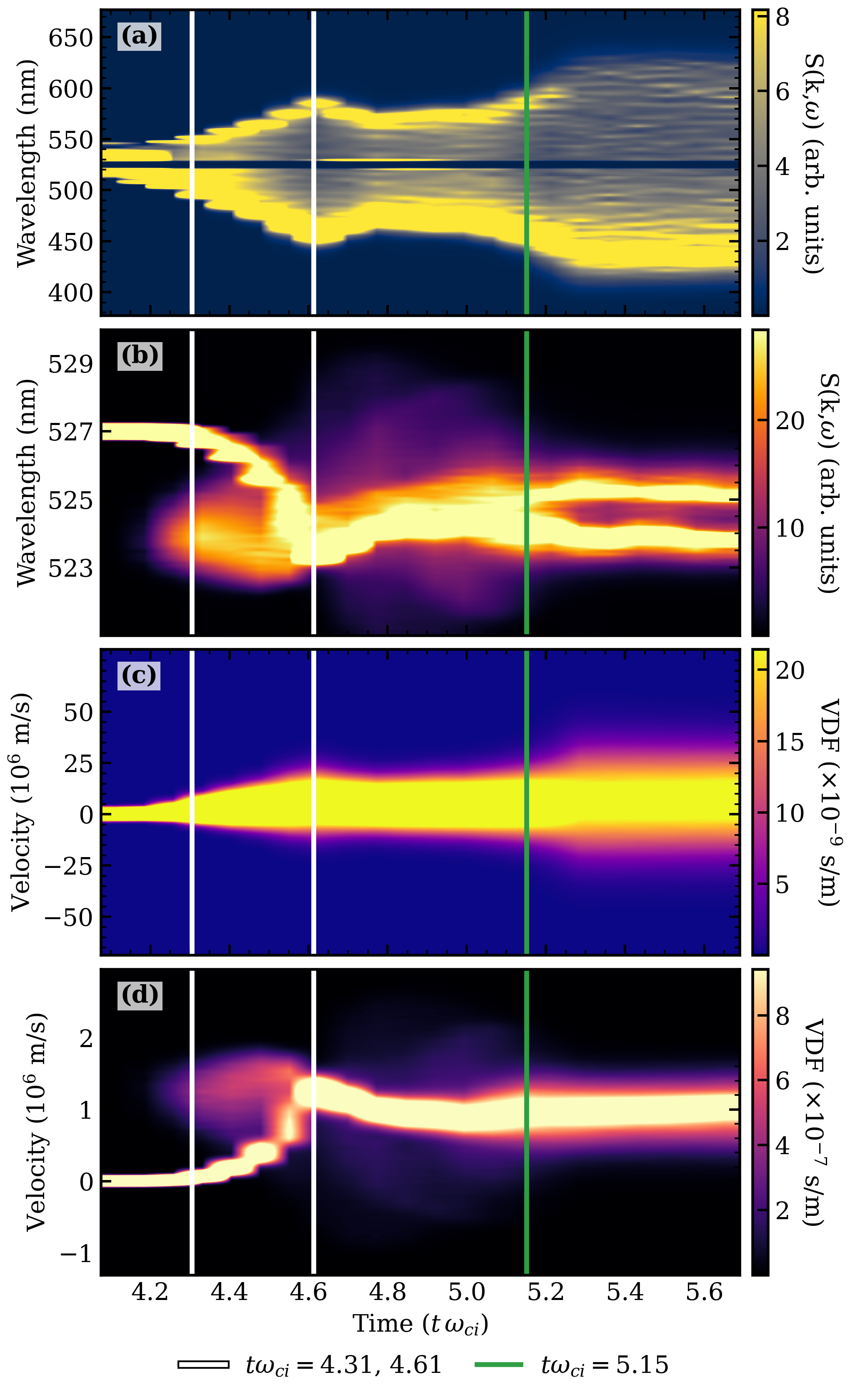}
    \hfill
    \caption{
    Ground truth PIC simulation spectra and VDFs in the analysis window. (a) EPW spectral feature; (b) IAW spectral feature; (c) electron VDF; (d) combined H/C ion VDF. Shock onset marked with white lines, carbon arrival marked with a green line.
    }
    \label{fig:ground_truth_spectra_vdfs}
\end{figure}

To mimic experimental data, spectra are extracted from the simulation at a fixed location ($5$ mm from the target, similar to experiments) in a streaked format (wavelength versus time) and separated into EPW and IAW images. A notch is also applied to the EPW spectra to block residual light near the probe wavelength, similar to experiments~\cite{katz2012}. The spectra and corresponding ground truth VDFs are shown in Figure~\ref{fig:ground_truth_spectra_vdfs}. The time window begins when moving plasma first enters the measurement location, and extends until the piston plasma has entered.  This interval was chosen as it contains both the onset of a collisionless shock composed primarily of H (visible through the rapid blue-shifting of the IAW feature from $t\omega_{ci}=4.31$ to $t\omega_{ci}=4.61$), and a multi-ion (CH) plasma as the shock downstream transitions to the piston (visible through the formation of two distinct spectral peaks around $t\omega_{ci}=5.15$). Thus, the effectiveness of the proposed basis representations can be tested in regions where non-equilibrium behavior and cross-species degeneracies are expected.

\subsection{Automatic Differentiation for VDF Optimization}
\label{sec:AD}

We implement a differentiable forward modeling framework for Thomson scattering that enables inference of VDFs from measured spectra. The method consists of a parameterized representation of the electron and ion VDFs, a forward model that maps these distributions to the spectral density $S(\mathbf{k}, \omega)$ and a gradient-based optimization process using automatic differentiation (AD)~\cite{baydin2018}. The forward model is implemented entirely with differentiable PyTorch operations~\cite{paszke2019}, so gradients propagate through every stage of the spectral calculation. At each iteration, a trial set of VDF parameters generates a model spectrum, which a loss function compares to the target spectrum. Gradients of the loss with respect to all model parameters are then computed through backpropagation and used to update the VDF parameters.

This approach is particularly efficient for high-dimensional parameter spaces: a single backward pass through the computational graph yields the gradient with respect to every parameter, at a cost comparable to one forward evaluation regardless of the number of parameters. This is a critical advantage over methods such as DE and MCMC, whose computational cost can grow rapidly with dimensionality. Parameter updates are performed with the Adam optimizer\cite{kingma2015} with default betas, $\beta_1=0.9$ and $\beta_2=0.999$.

The VDF parameters are optimized by minimizing the normalized mean squared error (NMSE) between the modeled and target spectra. For a spectrum sampled at $N_\lambda$ wavelength bins, the loss function is defined as

\begin{equation}
\mathcal{L}(\boldsymbol{\theta}) =
100 \times
\frac{
\frac{1}{N_\lambda} \sum_{j=1}^{N_\lambda}
\left[
S_{\mathrm{model}}(\lambda_j;\boldsymbol{\theta})
-
S_{\mathrm{target}}(\lambda_j)
\right]^2
}{
\mathrm{Var}\!\left(S_{\mathrm{target}}\right)
},
\end{equation}

\noindent where $\boldsymbol{\theta}$ denotes the set of VDF parameters and $\mathrm{Var}(S_{\mathrm{target}})$ is the variance of the target spectrum. This normalization makes the loss insensitive to the absolute amplitude of the spectrum and captures relative deviations in spectral structure. Fits were considered converged if the difference in NMSE between successive iterations was less than a threshold value; practical convergence criteria are detailed in Section~\ref{sec:fitting_procedure}. Gradients of $\mathcal{L}$ with respect to all model parameters are computed using AD in PyTorch and are used to iteratively update the VDF parameters.

\subsection{Basis Representation for VDF Reconstruction}
\label{sec:generalized_basis}

This work uses two basis sets in addition to the standard Maxwellian model as a control. The first is a superposition of Gaussian components, denoted Gaussian Mixture (GM). The second is a new generalized function that combines the super-Gaussian and kappa distributions into a single expression, denoted Super-Kappa (SK). The Maxwellian control is denoted MX.

GM involves a normalized superposition of $N_s$ Gaussian distributions for each species $s$ of the form
\begin{equation}
f_s(v) = \sum_{m=1}^{N_s} w_{s,m} \, f_G(v; \sigma^2_{s,m}, \mu_{s,m}),
\end{equation}
where $f_G(v; \sigma^2, \mu)$ is a Gaussian centered at mean $\mu$ with $1/e$ half-width $\sigma$. The weights $w_{s,m}$ satisfy $w_{s,m} \geq 0$ and $\sum_m w_{s,m} = 1$. When fitting with this technique, six Gaussian components were used to represent the electron VDF, while three components were used for each ion species. However, the number of components per species is flexible and can be adjusted if more or fewer non-thermal features are expected.

\begin{figure}[h]
    \centering
   \includegraphics[width=\columnwidth]{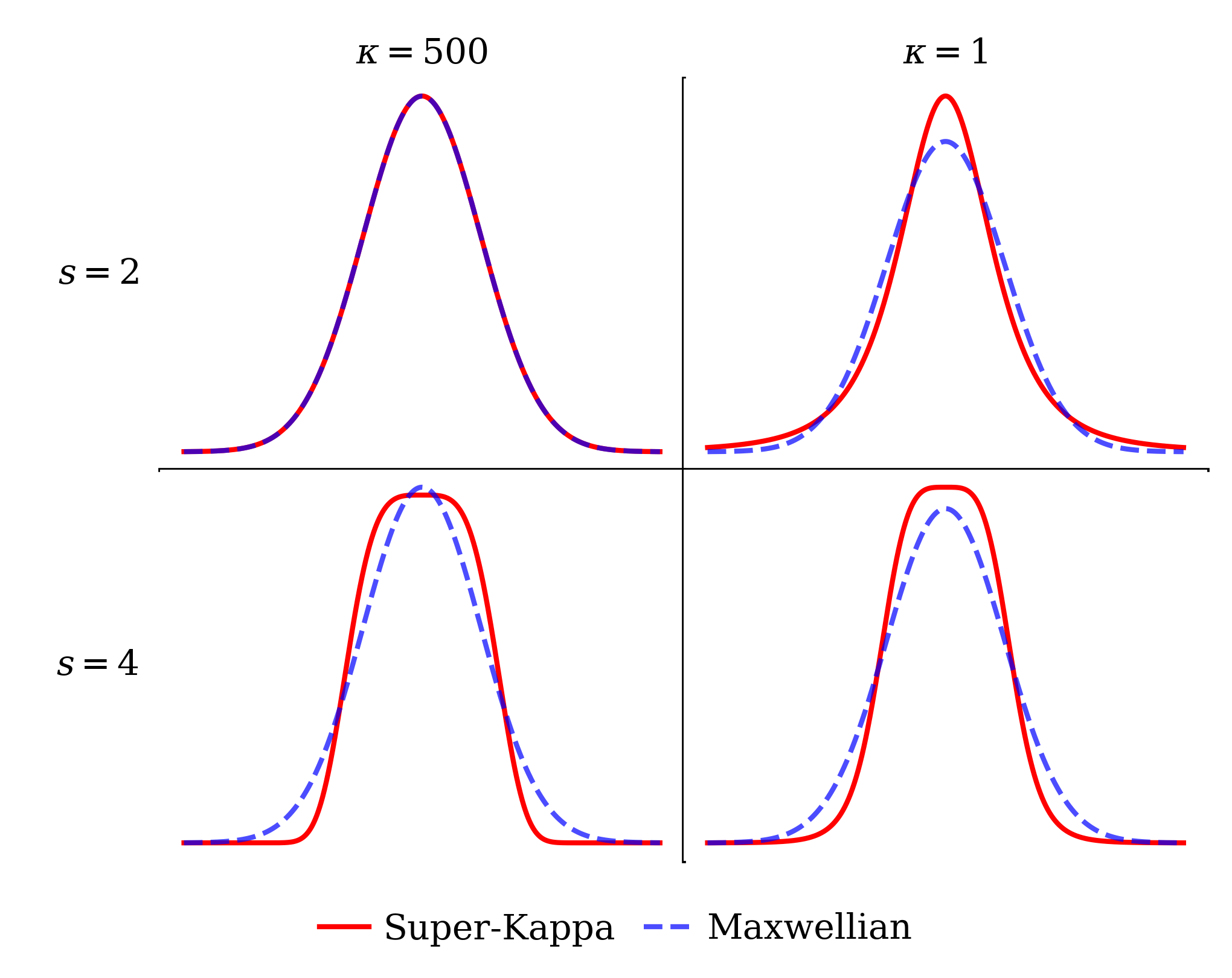}
    \caption{Effect of $s$ and $\kappa$ on SK distribution, with Maxwellian (blue) for comparison. SK reduces to Maxwellian for $s=2$, $\kappa\gg1$. Changing $s$ controls flatness, while changing $\kappa$ controls tail behavior.}
    \label{fig:sk_parameters}
\end{figure}

SK differs in principle from GM: rather than adding Gaussian components, it makes the underlying distribution itself more flexible while keeping it physically plausible. This matters, because preliminary non-physical bases for this work overfit badly and yielded unrealistic VDFs. The unnormalized SK distribution, which unifies the super-Gaussian~\cite{zheng1997,langdon1980} and kappa~\cite{vasyliunas1968,livadiotis2013} forms, is given by

\begin{equation}
    f(v) = \left[1 + \frac{1}{\kappa} \left(\frac{|v - \mu|}{\sigma} \right)^s \hspace{0.03in} \right] ^{-(\kappa + 1)}
    \label{eq:superkappellian}
\end{equation}

\noindent and is characterized by two new parameters $s$ and $\kappa$ in addition to the mean $\mu$ and $1/e$ half-width $\sigma$, which serve the same function as in a Gaussian. $s$ and $\kappa$ control the flatness and tail behavior of the distribution, respectively, and for $s=2$, $\kappa \gg 1$ the distribution reduces to a Gaussian (see Fig.~\ref{fig:sk_parameters}). This behavior makes SK an extension of MX in the case where just one SK distribution is used. In this work, one SK per population was used most frequently, except at times when large non-thermal effects were expected, such as near the shock onset; there a weighted sum of two SK distributions was used instead to offer additional flexibility.

We chose the SK basis for three reasons: 1) laser-heated plasmas form super-Gaussian distributions through preferential heating of slow electrons~\cite{langdon1980,supergaussian_laser_heated}; 2) kappa distributions are common in astrophysical plasmas and have been applied to laboratory and fusion plasmas~\cite{kappa_dist_1,kappa_dist_2,livadiotis2013}; and 3) collisionless shocks are expected to produce non-Maxwellian features~\cite{feldman1983,sckopke1990}, making non-Maxwellian basis functions a natural choice.

The MX control followed the standard convention of a single Maxwellian (Gaussian) per electron and ion population with temperature $T$ and drift velocity $v_d$. This naturally allows the physical state of the plasma to be read off from the optimized $T$ and $v_d$. We emphasize that the alternative basis choices remove the physical interpretability of the fit parameters. Using multiple components to represent a single VDF also introduces degeneracies, as multiple parameter combinations will reproduce the same VDF and spectrum; information about the system must be obtained from the VDF itself instead of the fitted parameters. Optimization for all methods was performed in a re-parameterized space, which keeps the gradients of parameters with very different scales comparable in magnitude and enforces physical constraints (temperatures remain positive and ion species fractions sum to unity).

\subsection{Fitting Procedures}
\label{sec:fitting_procedure}

The electrons and ions were fit in a two-stage process, following standard TS practice~\cite{froula2006,follett2016}. The electron feature, being largely independent of the ion state, was fit first to obtain the electron density and VDF. The ion feature is highly dependent on the electron state and was thus fit afterwards with the electron fit held fixed to reduce the number of free parameters. The ground truth electron feature is largely consistent with a single electron population and was modeled as such. The ion feature and the experimental setup indicated three ion populations: cold ambient hydrogen at first, followed by hot hydrogen and carbon in the expanding piston.

The convergence threshold for both the electron and ion stage was defined to be $10^{-6}$, such that $|\mathcal{L}_i - \mathcal{L}_{i-1}| \leq 10^{-6}$ was considered to be a converged fit. An additional iteration hard limit was imposed both to save time and to minimize overfitting. For the electron stage, which was generally easier to initialize and had fewer free parameters, fits were stopped after 1500 iterations, which was found to be enough for nearly all fits provided a reasonable initialization. The few fits that did not converge within this limit were generally found to be oscillating about the loss minimum while avoiding the $\delta\mathcal{L}$ threshold; the effects of additional iterations were negligible for these fits. The ion stage had a hard limit of 4000 iterations, settled on after considerable trial and error: a fit that converged to a reasonable solution at all typically did so within 4000 iterations. These limits are a practical choice rather than a requirement: they prevent oscillating or over-fitted solutions from running indefinitely while leaving the fit quality of capped solutions largely unchanged.

The electron and ion spectra provide substantial information to guide initialization. In general, spectral peaks can be aligned relatively easily by adjusting the mean/drift velocity of each basis component. The formation of a second distinct resonance peak in the ion spectrum around $t\omega_{ci}\approx5.15$ is indicative of heavy carbon entering the system, which informs the initial ion fractions---a crucial step for multi-ion reconstruction, because of the large degeneracy addressed in Section~\ref{sec:uncertainty}.

Initialization is a non-trivial part of the optimization process, and each fit is highly sensitive to it: the initial guess must capture the core spectral features well and then let the basis flexibility take over during optimization. In practice, this led to using the Maxwellian model to obtain a ``close enough'' initialization and then injecting the new basis choices---for GM, lower-weighted components were added near the core Maxwellian to account for asymmetries and non-thermal tails, whereas for SK the existence of the additional $s$ and $\kappa$ parameters (which were initialized at 2 and 100, respectively, to mimic a Maxwellian) naturally allowed the original Maxwellian to change shape during the optimization. Because the MX group was initialized with just the core Maxwellian shared by GM and SK, differences between the methods are attributable to the basis, and any bias introduced by the initialization is common to all methods.

Physical intuition is also required to constrain the potential solutions. These constraints were applied both before and after the fit to improve the initialization and discard unphysical solutions. For example, obvious fitting artifacts, unrealistically broad low-amplitude ion populations, and large discontinuities in the VDFs or species fractions between consecutive time steps were regarded as unphysical and therefore rejected. The expectation of a roughly continuous evolution of the VDFs and plasma parameters was also used to initialize subsequent fits with similar parameters to the previous fit.

\section{Results and Discussion}
\subsection{Spectral Reconstruction}
\label{sec:spectral_reconstruction}

\begin{figure}[!htbp]
    \centering
    \includegraphics[width=\columnwidth]{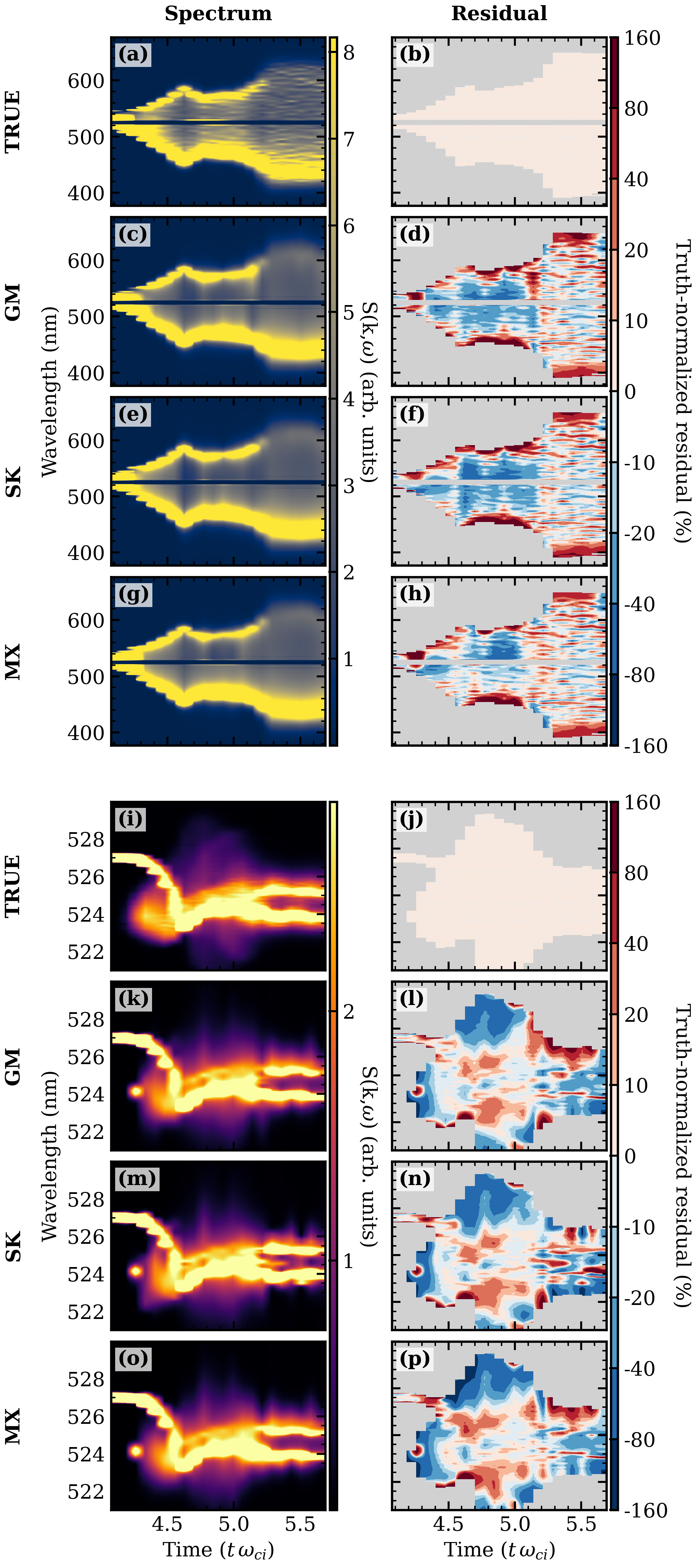}
    \hfill
    \caption{
    Reconstructed EPW and IAW spectral features (left) and normalized residual maps (right) for each basis method, with ground truth for comparison. (a) Ground truth EPW spectrum; (c)/(d) GM EPW spectrum/residuals; (e)/(f) SK EPW spectrum/residuals; (g)/(h) MX EPW spectrum/residuals; (i) Ground truth IAW spectrum; (k)/(l) GM IAW spectrum/residuals; (m)/(n) SK IAW spectrum/residuals; (o)/(p) MX IAW spectrum/residuals. Ground truth spectra are plotted prior to Gaussian noise injection.
    }
    \label{fig:spectral_reconstruction}
\end{figure}

We first evaluate the ability of each proposed basis representation to reconstruct the ground truth spectra. Figure~\ref{fig:spectral_reconstruction} presents the best-fit EPW and IAW spectra obtained using each basis, along with corresponding relative error maps computed with respect to the ground truth. The reconstructed spectra are compared directly to the true noise-free simulation data, with the error maps visualizing the residuals. The pointwise residuals $R_i$ in this and subsequent figures are normalized by the true magnitude:

\begin{equation}
    R_i = \frac{\text{fit}_i - \text{true}_i}{\text{true}_i},
\end{equation}

\noindent where true indicates the noise-free spectrum or VDF. Points where the true value falls below 1\% of its maximum at that time step are excluded to avoid disproportionately large relative errors where the signal is small. The residual maps are also colored by discrete, nonuniform bins to highlight contrast and structure. Overall, the reconstructed EPW and IAW spectra reproduce the dominant characteristics of the system for all basis choices, including peak locations, relative amplitudes and general temporal evolution.

The EPW residual structure displays a clear negative residual pattern surrounding the notch at earlier times, and is noticeably noisier than the IAW structure throughout. We attribute the negative residuals to the low amplitude of this region, which both amplifies small absolute differences under the pointwise normalization and, together with the notch, leaves the region less constrained than the spectral peaks. The noise, which can be seen in the ground truth EPW spectrum, is an artifact of the forward model, where the small amount of noise in the underlying VDF is amplified by the low particle count near the EPW phase velocity.

The IAW residual structure displays qualitatively different behavior---it is not uniformly distributed across wavelength and displays sign variation that appears mostly random, ruling out any major systematic over/underestimation in the fitted spectra. A key finding from the residual map is its similar structure across the three methods, such as the large negative residual around $\lambda = 528$ nm and $t\omega_{ci}\approx4.69$--5.00, indicating that regions of high error are more likely a product of artifacts in the target data or the injected noise instead of a basis-specific defect. This could be attributed in part to the pre-processing of the underlying simulation VDFs, which were smoothed using a Savitzky-Golay filter~\cite{savitzky1964} to reduce statistical noise, providing a potential source of non-physical artifacts.

Some notable differences between the fitted and true IAW spectra are the bright spot around $\lambda = 524$ nm at $t\omega_{ci}=4.23$ and the appearance of a small third peak around $t\omega_{ci}=5.15$ for SK, both of which are attributed to noise. The $t\omega_{ci}=4.23$ artifact is due to noise obscuring the small signal at that location and causing the model to overestimate its amplitude. The third peak is a localized overfitting artifact, as SK's additional flexibility lets it fit a noise feature that MX cannot represent. Despite also having much greater flexibility, GM does not display this artifact as clearly, which we attribute to the different shapes each basis can create. However, the artifact remains small and does not greatly affect the reconstructed VDF or moments (Sec.~\ref{sec:time_resolved_parameters}).

Error variation and magnitude in the IAW residual map also increase after carbon enters the system near $t\omega_{ci}=5.15$. This is a preliminary indicator of the increased fit degeneracy and complexity that is associated with a multi-ion plasma, which will be discussed further in Section~\ref{sec:uncertainty}.

\subsection{VDF Reconstruction}
\label{sec:vdf_reconstruction}

\begin{figure}[!htbp]
    \centering
    \includegraphics[width=\columnwidth]{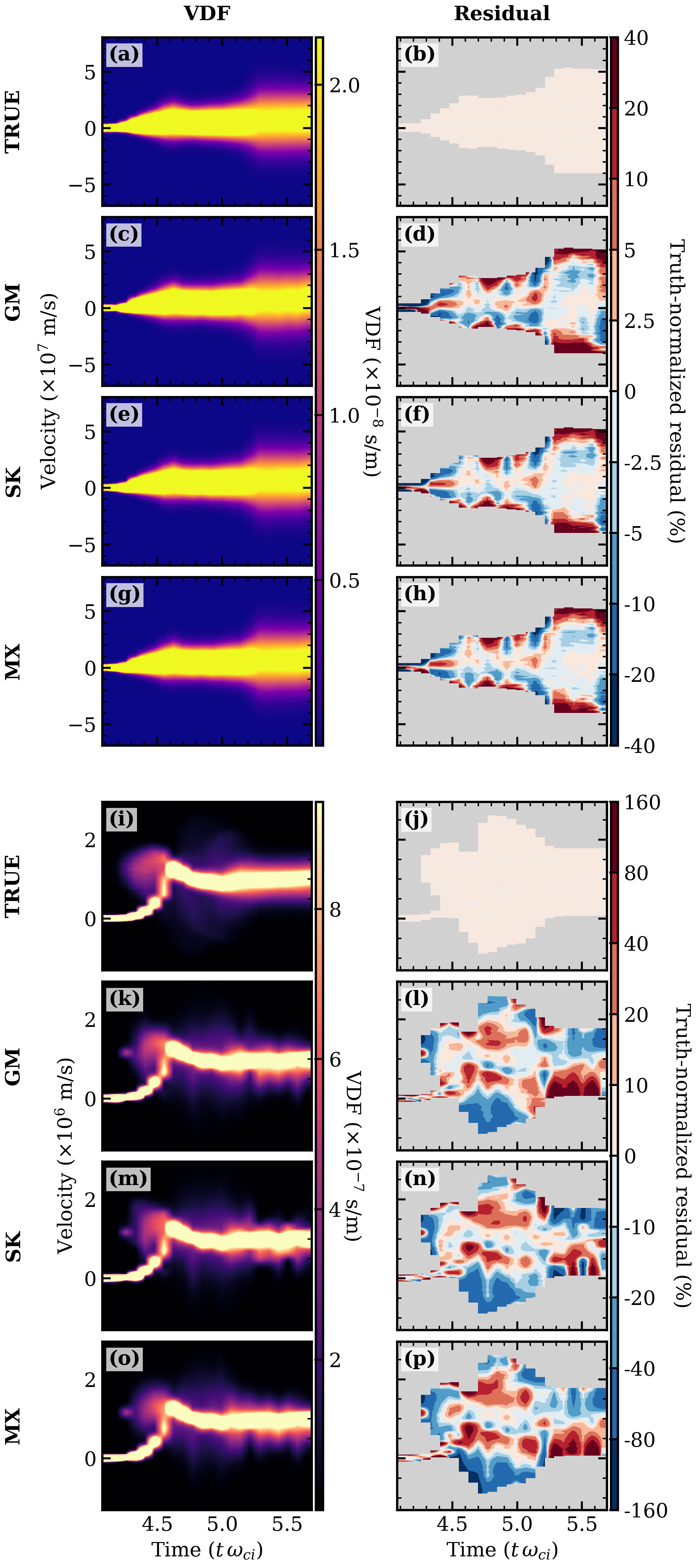}
    \hfill
    \caption{
    Reconstructed electron and ion VDFs (left) and normalized residual maps (right) for each basis method, with ground truth for comparison. (a) Ground truth electron VDF; (c)/(d) GM electron VDF/residuals; (e)/(f) SK electron VDF/residuals; (g)/(h) MX electron VDF/residuals; (i) Ground truth ion VDF; (k)/(l) GM ion VDF/residuals; (m)/(n) SK ion VDF/residuals; (o)/(p) MX ion VDF/residuals.
    }
    \label{fig:vdf_reconstruction}
\end{figure}

While the spectral reconstruction shows strong overall agreement with the ground truth, this does not guarantee accurate reconstruction in phase space as distinct VDFs can produce nearly identical spectra. To better assess the fit quality, we extend the analysis to phase space, with the reconstructed electron and ion VDFs plotted in Figure~\ref{fig:vdf_reconstruction}. We find that the general shape and time evolution of both electron and ion VDFs match the ground truth VDFs closely in Figure~\ref{fig:ground_truth_spectra_vdfs} for all basis choices.

The residual maps in Figure~\ref{fig:vdf_reconstruction} are again normalized by the true magnitude and compared to the ground truth VDF. Their structure highlights a few key characteristics of the phase-space reconstruction. First, the electron reconstruction is accurate for all basis methods: the residuals are low in magnitude and show no systematic sign. Second, the ion fits are substantially more difficult, with much larger residuals across both velocity and time.

The ion evolution can be split into three main regions: the immediate upstream ($t\omega_{ci}<4.61$), immediate downstream ($4.61 \leq t\omega_{ci} < 5.15$), and carbon-dominated ($t\omega_{ci}\geq5.15$). The immediate upstream, which includes the shock front and foreshock, delivers most of the benefit from the new basis choices: the average normalized residual magnitude there is 20.3\%, 20.5\%, and 38.7\% for GM, SK, and MX, respectively.

The immediate downstream displays little to zero noticeable improvement for GM and SK over MX, instead showing very similar residual structure across all methods, with an average correlation of 95.1\%. This indicates one or both of the following possibilities: 1) the additional basis flexibility provided by GM and SK is being under-utilized due to the complexities of the noisy inversion process or subpar initializations, or 2) assuming a single Maxwellian distribution for each population in this region (in this case, the initial ambient hydrogen and the piston hydrogen were still being modeled as two distinct populations after they collided) is sufficient to recover a reasonable reconstruction of the true VDF.

The carbon-dominated region demonstrates a few points of interest. First, all bases tend to overestimate the VDF at lower velocities and underestimate it at higher velocities; that ubiquity points to the spectral inversion as the cause. Second, GM and MX show similar structures, correlating at 86.7\% in this region, but SK does not---a difference we attribute to the shapes each basis can build: GM is made of Gaussians and so struggles to produce forms such as a flat-top distribution, which SK captures readily.

\begin{figure}[!htbp]
    \centering
    \includegraphics[width=\columnwidth]{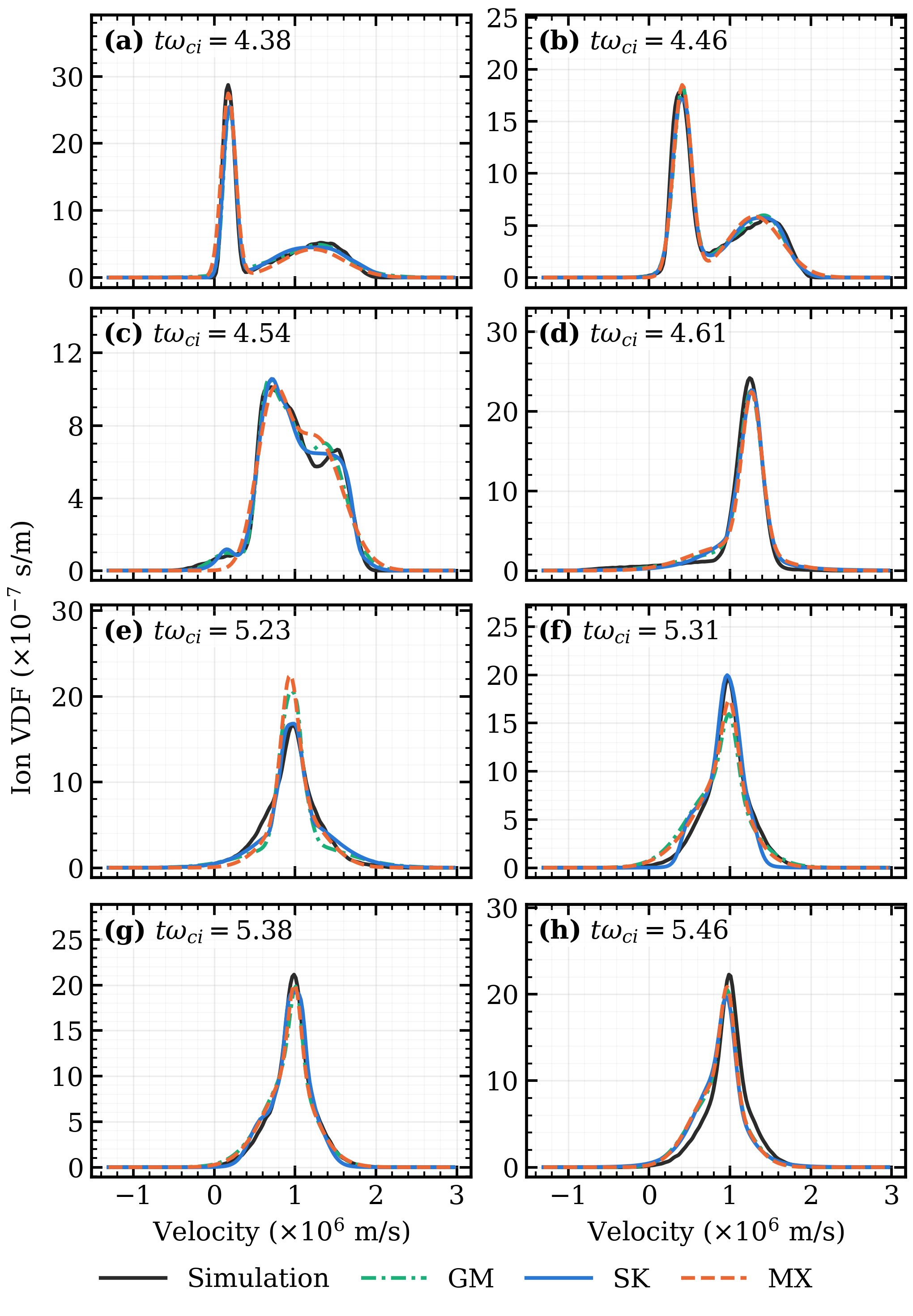}
\caption{Ion VDFs for specific times. (a)-(d) correspond to times in the immediate vicinity of the shock; (e)-(h) correspond to times after carbon has entered the scattering volume. GM and SK display an improvement in reconstruction quality over MX for (a)-(c), most substantial in (c). At later times the methods are broadly comparable, although SK more closely reproduces the peak height in (e) and (f).}
\label{fig:1d_ion_vdf_grid}
\end{figure}

While the overall temporal evolution of the reconstructed VDF gives high-level insight into the differences between basis choices, individual velocity lineouts in time better highlight the specific advantages (or lack thereof). This is shown in Figure~\ref{fig:1d_ion_vdf_grid}. These times were selected as they represent the two regions of highest ion fit complexity in the window of analysis: the first region, $t\omega_{ci}=4.38$--$4.61$, is at the boundary of the shock and demonstrates significant deviations from a thermal distribution, whereas the second region, $t\omega_{ci}=5.23$--$5.46$, is when carbon has noticeably entered the system and greatly increased the fit degeneracy.

The individual ion VDFs in Figure~\ref{fig:1d_ion_vdf_grid} show improved reconstruction quality in the immediate upstream, $t\omega_{ci}=4.38$--$4.54$. At $t\omega_{ci}=4.38$ and $t\omega_{ci}=4.46$, the added flexibility mainly captures the shape of the non-thermal shocked plasma. The differences between the bases are most pronounced at $t\omega_{ci}=4.54$, where MX is fully unable to capture the true shape, whereas GM and SK are much closer to the ground truth. GM best captures the skewness of the piston population at $t\omega_{ci}=4.46$, since three Gaussian components can reconstruct asymmetric VDFs~\cite{ross2014}. A skewness parameter was considered for SK but was not pursued in the interest of simplicity (see Section~\ref{sec:conclusion}).

Immediately after the shock ($t\omega_{ci}=4.61$) and at the latest times ($t\omega_{ci}=5.38$--$5.46$), the methods perform comparably, with GM and MX very similar throughout, consistent with the residual maps in Figure~\ref{fig:vdf_reconstruction}. At $t\omega_{ci}=5.23$ and $5.31$, however, SK reproduces the peak of the ion VDF noticeably better: GM and MX overshoot it at $5.23$ and undershoot it at $5.31$, while the wings are reproduced similarly by all methods. We note that all methods were able to reconstruct the high-velocity non-thermal tails present in the immediate downstream of the shock, corresponding to rapid cyclotron motion of the shocked ions~\cite{sckopke1990}. This is somewhat surprising for MX, and suggests that modeling the shocked hydrogen ions with a bi-Maxwellian may be sufficient to reconstruct these high-velocity tails.

\begin{figure}[!htbp]
    \centering
    \includegraphics[width=\columnwidth]{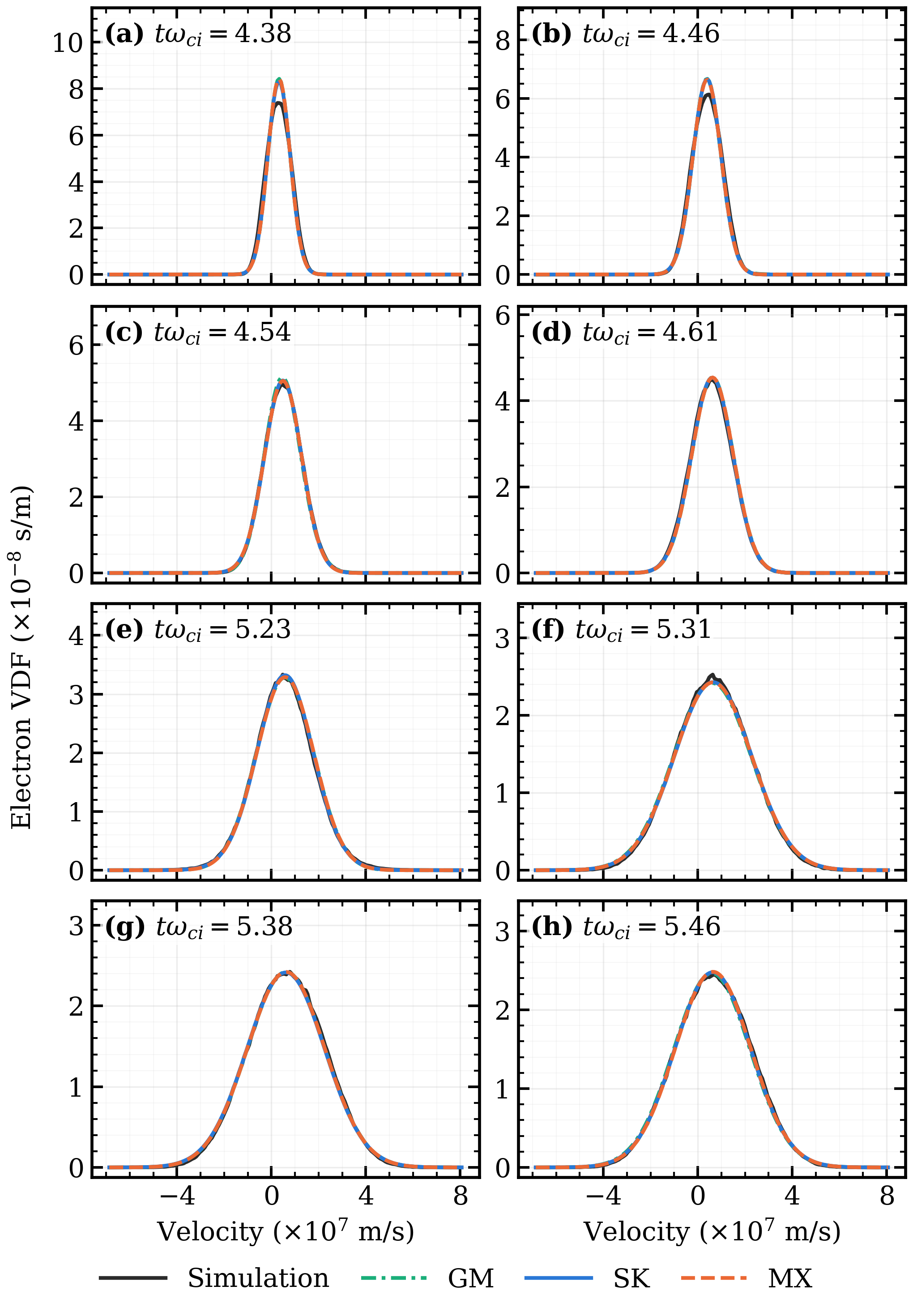}
\caption{Electron VDFs for specific times. (a)-(d) correspond to times in the close vicinity of the shock. (e)-(h) correspond to times after carbon has entered the scattering volume. Electrons are most non-thermal in the immediate upstream and almost completely Maxwellian at later times. All methods agree closely with each other for all times considered.}
\label{fig:1d_electron_vdf_grid}
\end{figure}

The electrons behave differently from the ions over this interval. Their lighter mass lets them equilibrate on time scales much shorter than the shock evolution; the electron--electron collision frequency was determined to be $\nu_{ee} \gtrsim 60~\omega_{ci}$ throughout the window~\cite{nrlformulary}. The electrons therefore display only slight non-thermal behavior, if any. Figure~\ref{fig:1d_electron_vdf_grid} displays temporal lineouts of the electron VDFs. As MX uses only one Maxwellian for the electron population, agreement between MX and the ground truth indicates that the electrons are fully thermal at that time. The agreement of all methods in the slight non-thermal regions suggests that the non-thermal effects on the EPW spectrum are negligible within the 10\% injected noise.

Overall, both new basis methods provide a meaningful improvement in the reconstructed ion VDFs in the times preceding the shock, while demonstrating largely comparable performances at times after the shock. This aligns with the expectation that the largest improvements would appear where the non-thermal deviations are largest. MX was nonetheless surprising in the immediate downstream, fitting a non-thermal VDF comparably to GM and SK with a substantially simpler model. We attribute this to the VDF shape in this region: while the overall VDF is certainly non-thermal, it can be modeled well enough by treating it as a mixture of two distinct, thermal hydrogen populations, up to the spectral noise and general obfuscation of the forward model inversion.

SK produced the clearest overfitting artifact, the spurious spectral peak around $t\omega_{ci}=5.15$. Even so, the similar performance of all methods at later times indicates that neither GM nor SK overfit substantially despite their additional degrees of freedom. Overfitting nevertheless remains a concern, since too much flexibility often drives the optimizer to non-physical solutions.

\subsection{Time-Resolved Plasma Parameters}
\label{sec:time_resolved_parameters}

In addition to the reconstructed VDFs, physical moments provide a secondary, coarser metric of fit quality, and a more physically interpretable one: they track the time evolution of density, flow velocity, and temperature for individual electron and ion populations.

In all cases, the optimized VDFs are normalized such that:

\begin{equation}
\int_{-\infty}^{\infty} f(v)\, \mathrm{d}v = 1
\end{equation}

As a result, the number density of a given population is not obtained from the zeroth velocity moment; rather, the electron density is stored as a separate fit parameter and optimized independently, which in turn fixes the ion population densities via the quasineutrality assumption. Using normalized VDFs allows for the first and second moments to be directly integrated, with species flow velocity and temperature being defined as:

\begin{equation}
    u_s = \int_{-\infty}^{\infty} v\, f_s(v)\, \mathrm{d}v,
\end{equation}
\begin{equation}
    T_s = \frac{m_s}{e} \int_{-\infty}^{\infty} (v - u_s)^2 f_s(v)\ \mathrm{d}v,
\end{equation}
where $m_s$ is the mass of species $s$ and $e$ is the elementary charge. Note that this is the one-dimensional temperature along $\mathbf{\hat{k}} = \hat{z}$.

\begin{figure}[t!]
    \centering
    \includegraphics[width=\columnwidth]{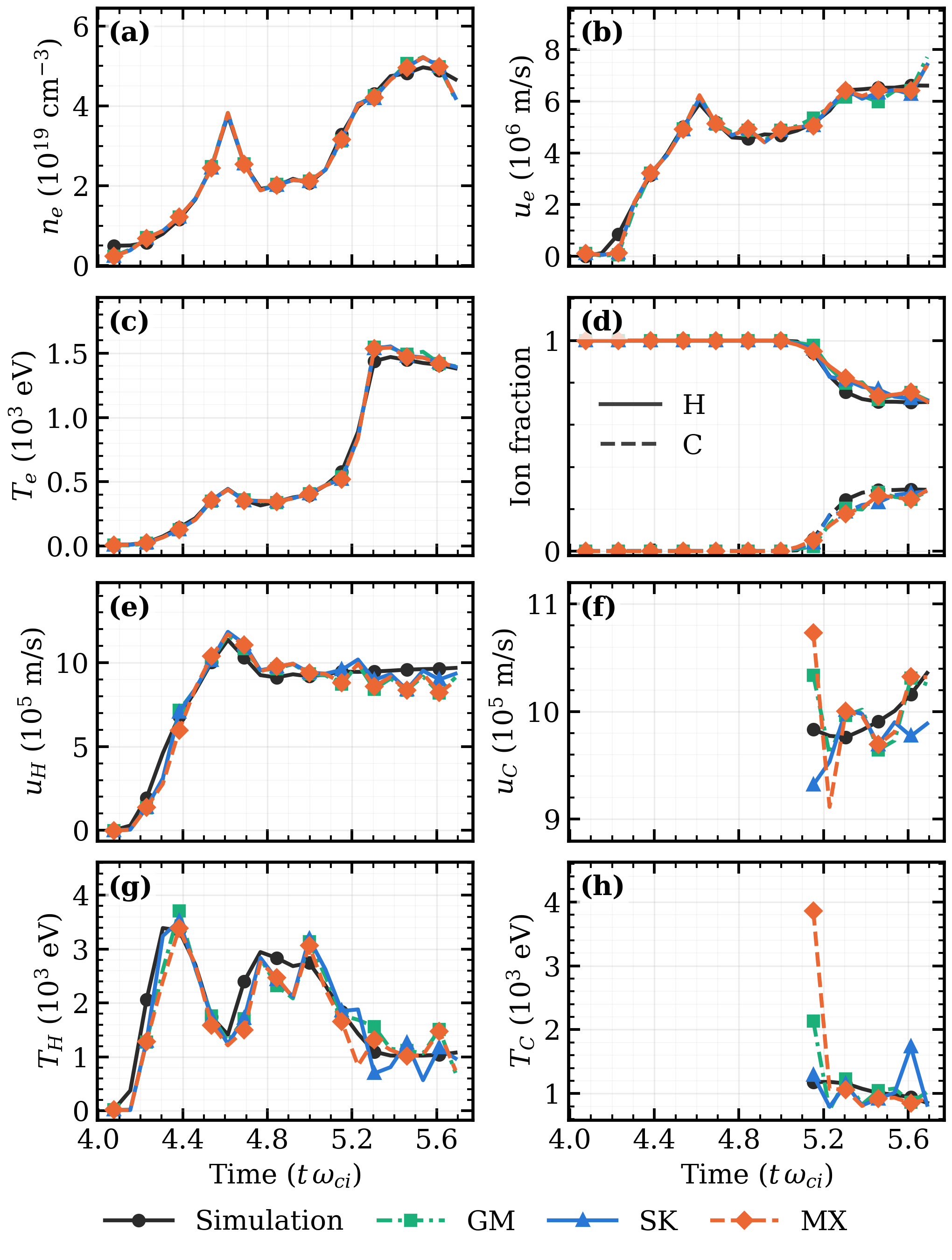}
    \caption{
        Time-resolved plasma moments computed from the fitted distributions for each basis, with simulation ground truth for reference. (a) electron density; (b) electron flow velocity; (c) electron temperature; (d) ion relative weights; (e) hydrogen flow velocity; (f) carbon flow velocity; (g) hydrogen temperature; (h) carbon temperature. Carbon moments start at $t\omega_{ci}=5.15$ as the carbon fraction is zero or negligible prior to this time.
    }
    \label{fig:moments_vs_time}
\end{figure}

The temporal evolution of all computed moments is displayed in Figure~\ref{fig:moments_vs_time}. The ion number density $n_j$ is determined from the relative ion weights and the quasineutrality assumption $n_e = \sum_j Z_j n_j$, where $Z_j$ is the charge state of $j^{th}$ ion species (assumed to be 1 and 6 for hydrogen and carbon, respectively). The ion weights $w_j$ are defined such that $n_j = w_j n_i$, where $n_i$ is the total ion density. Combining these gives the number density of the $j^{th}$ ion species as:

\begin{equation}
    n_j = \frac{w_j n_e}{\sum_{j'} Z_{j'} w_{j'}}
\end{equation}

The electron moment reconstruction is accurate, with the median percent difference from the true moments between 2--5\% for all moments across all methods. This is independent evidence for the quality of the electron reconstruction: because the density and the normalized VDF are fit separately, VDF shape alone does not guarantee a good density recovery. Since the ion stage depends strongly on the electron stage, accurate electron VDFs and moments also confirm that the ion fit is not substantially biased by a poor electron solution.

The ion moment reconstruction tracks the trends observed in the VDF reconstruction stage, with basis-specific improvements in the immediate upstream, little difference in the immediate downstream, and generally noisier moment reconstruction in the carbon region. The upstream improvements over MX are smaller in moment space than in phase space, because integrating over the VDF discards the finer shape details. As expected, the reconstructed ion moments are lower quality on average than the electron moments, as the ion fit is inherently more complex due to the existence of multiple populations.

In particular, the ion flow velocity is captured more accurately than the ion temperature, with a median percent error across all methods of 2--5\% compared to 10--17\%. We attribute this to the direct effect of flow velocity on the spectral peak locations, which makes it comparatively easy to optimize, whereas the subtler signature of temperature is often masked by the inversion and by noise.

We also note that the optimized ion fractions are highly dependent on the initialization. The presence of a large degeneracy between different ion species, discussed in Section~\ref{sec:uncertainty}, causes the optimizer to struggle with determining the relative weights of each population. To achieve a reasonable reconstruction, the carbon onset can usually be identified from the spectral structure---the appearance of additional peaks~\cite{fried1971,glenzer1996}---with the expectation of a smooth increase in the carbon fraction guiding subsequent initializations.

\section{Uncertainty Analysis}
\label{sec:uncertainty}

Because the inverse problem is complex and ill-posed~\cite{tarantola2005}, numerous sets of optimized parameters and VDFs yield reasonable solutions, especially with a noisy target spectrum. Each fit is also highly sensitive to an initialization that is physically motivated yet still subjective. We therefore use a Monte Carlo (MC) approach to probe the basin of attraction of the best fit. Many initializations converging to a similar solution indicate a well-constrained fit, whereas slight deviations in the initial guess that lead to very different results indicate a poorly constrained one.

The MC approach involved taking the parameter vector $\theta$ that initialized the best fit, applying a random perturbation to each component $\theta_i$, and refitting the perturbed parameter vector. This was performed 150 times to obtain a sufficiently large sample size. Each perturbation $\delta\theta_i$ was chosen at random and capped at 30\% of the initial parameter, such that $|\delta\theta_i| \leq 0.3\theta_i$. This cap was chosen to ensure most of the perturbed initializations did not stray too far from the base initialization, as this would make them largely nonphysical in the context of the true spectrum, while still exploring a wide range of initializations.

This analysis was reserved for the ion fitting stage of two key times, $t\omega_{ci}=4.54$ and $t\omega_{ci}=5.31$. These times were chosen as they are representative of their respective regions---$t\omega_{ci}=4.54$ of the non-thermal VDFs in the immediate upstream and shock regions; $t\omega_{ci}=5.31$ of the period after carbon has substantially entered the scattering volume. The electron stage was not analyzed given the overall high quality of the reconstructed moments and VDFs at all times.

The MC samples revealed additional structures and degeneracies that required more careful handling than naively pooling the entire solution set. Many of the refit moments formed multi-modal distributions while having similar NMSE to the initial best fit, indicative of multiple local minima in the loss function. To better understand the solution landscape and uncertainties in particular solutions, the full samples were fit with a Gaussian Mixture Model (GMM, not to be confused with the basis method GM) of $k$ Gaussians~\cite{dempster1977,pedregosa2011}, where $k$ is the smallest value near the minimum of the Bayesian Information Criterion (BIC)~\cite{schwarz1978}, defined as

\begin{equation}
\text{BIC}(k) = -2\,\ln L(k) + p(k)\ln n.
\end{equation}

\noindent Here $n$ is the number of samples, $k$ the number of Gaussian components, and $L(k)$ is the likelihood function, representing the probability that the fitted $k$-component mixture matches the data. $p(k)$ is the number of free parameters: $k-1$ mixture weights (they must sum to 1), plus each component's own mean and covariance. This second term penalizes high-$k$ models to minimize overfitting, prioritizing the simplest possible model that gives a strong fit. The samples were primarily bimodal ($k=2$), though a smaller subset of points occasionally formed a third group. No third group held enough points to be meaningful, so this analysis considers only the two largest groups.

Once the majority groups had been identified, the best fit at that time was sorted into one of the two groups to identify the ``primary'' group; ``secondary'' denotes the other. Group statistics can then be isolated and compared, rather than mixing two distinct populations by sampling over the entire set~\cite{stephens2000}.

The GMM fit was performed on a moment or pair of moments for each given fit, as this forms a simple yet physically meaningful representation of the full sample. For example, at $t\omega_{ci}=4.54$, the only ion present in the scattering volume is hydrogen, so the only relevant moments are the hydrogen temperature and flow velocity. Temperature was chosen as the fitted moment for this time because it reduced the uncertainty in the primary groups for all bases, unlike splitting by flow velocity.

\begin{figure}[t]
    \centering
    \includegraphics[width=\columnwidth]{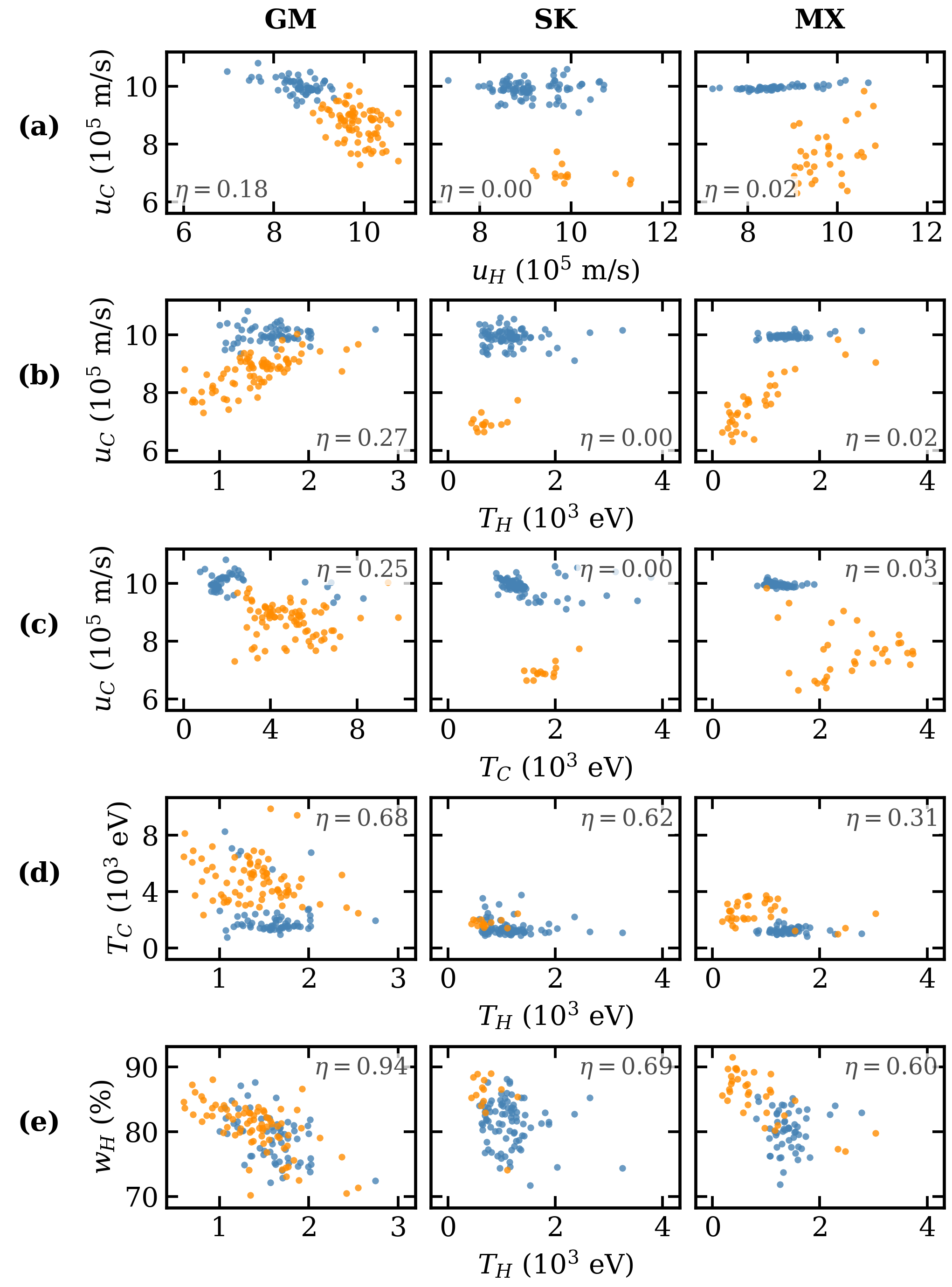}
    \caption{
        Monte Carlo samples for different moment pairs. (a) $u_C$ vs. $u_H$; (b) $u_C$ vs. $T_H$; (c) $u_C$ vs. $T_C$; (d) $T_C$ vs. $T_H$; (e) $w_H$ vs. $T_H$. Columns correspond to basis. Colors are determined by the GMM fit on (a). $\eta$ is the overlap between the two groups: 0 means the groups are completely distinct, 1 means the groups are identical. Noticeable splitting is observed in (b), (c), and (d), indicating that the grouping found in (a) remains distinct for other moment pairs involving carbon. (e) is shown as a negative control for a hydrogen-only pair, displaying minimal splitting.
        }
        \label{fig:basis_comparison_correlation_grid}
    \end{figure}

For $t\omega_{ci}=5.31$, a different mechanism was used to better reflect the fit's inherent grouping. As alluded to in earlier sections, the presence of carbon introduces a notable H/C degeneracy---unsurprising, given that a two-ion plasma supports two ion-acoustic modes with species-dependent damping~\cite{fried1971,glenzer1996,williams1995}---initially noticed through GMM fitting to pairs of moments. Figure~\ref{fig:basis_comparison_correlation_grid} shows evidence of this: a GMM fit was performed on row (a) and identified a clear split for each basis. This exact grouping is plotted in the subsequent rows and demonstrates a meaningful split in most moment pairs that involve both hydrogen and carbon while showing a much weaker split for row (e), which involves hydrogen moments only. This suggests the presence of two distinct solution groups that involve an interchange between hydrogen and carbon moments while being relatively agnostic to the basis choice, indicating that this is a fundamental property of the multi-ion fit itself and not a basis-specific artifact. The Bhattacharyya coefficient\cite{kailath1967} $\eta$ between the two groups is labeled for each moment pair, where 0 indicates full separation and 1 means that the groups are identical.

The consequences of this degeneracy will be discussed in more detail shortly---for now, it provides a meaningful way to determine the GMM grouping: for each basis, a GMM fit was performed on every moment pair and compared to every other moment pair via the Adjusted Rand Index (ARI)~\cite{rand1971,hubert1985}, which is defined as

\begin{equation}
    \text{ARI} = \frac{R - \mathbb{E}[R]}{R_{\max} - \mathbb{E}[R]},
\end{equation}

\noindent where $R$ is the fraction of item pairs that two partitions agree on (placed in the same group by both, or in different groups by both), $\mathbb{E}[R]$ is the value $R$ would take on average if the two partitions were unrelated/random, and $R_{\max}=1$ is the value for two identical partitions. ARI $=1$ means the partitions match exactly; ARI $\approx 0$ means they agree no better than chance. The pair with the highest average ARI to all other pairs, that is, the pair that demonstrates the most consistent grouping across all other pairs, was chosen to form the solution groups. For example, take the GM column in Figure~\ref{fig:basis_comparison_correlation_grid}. Comparing the flow velocity pair in row (a), $u_C$ vs. $u_H$, to the moment pairs in rows (b), (c), (d), and (e) yields an ARI of 0.76, 0.32, 0.41, and exactly zero, respectively. Thus, considering only these five pairs, pair (a) has an average ARI with the others of 0.37. We repeat this for each pair and select the highest average to form the solution groups. In this way, the solution groups for each basis were chosen to be the most distinct groups on average for all possible pairs of moments, ensuring they best represent the actual bimodal solution structure. Note that this chosen pair is not necessarily the same as the reference pair in Figure~\ref{fig:basis_comparison_correlation_grid} (a), which was fixed for all bases for visualization purposes.

\begin{figure}[!htbp]
    \centering
    \includegraphics[width=\columnwidth]{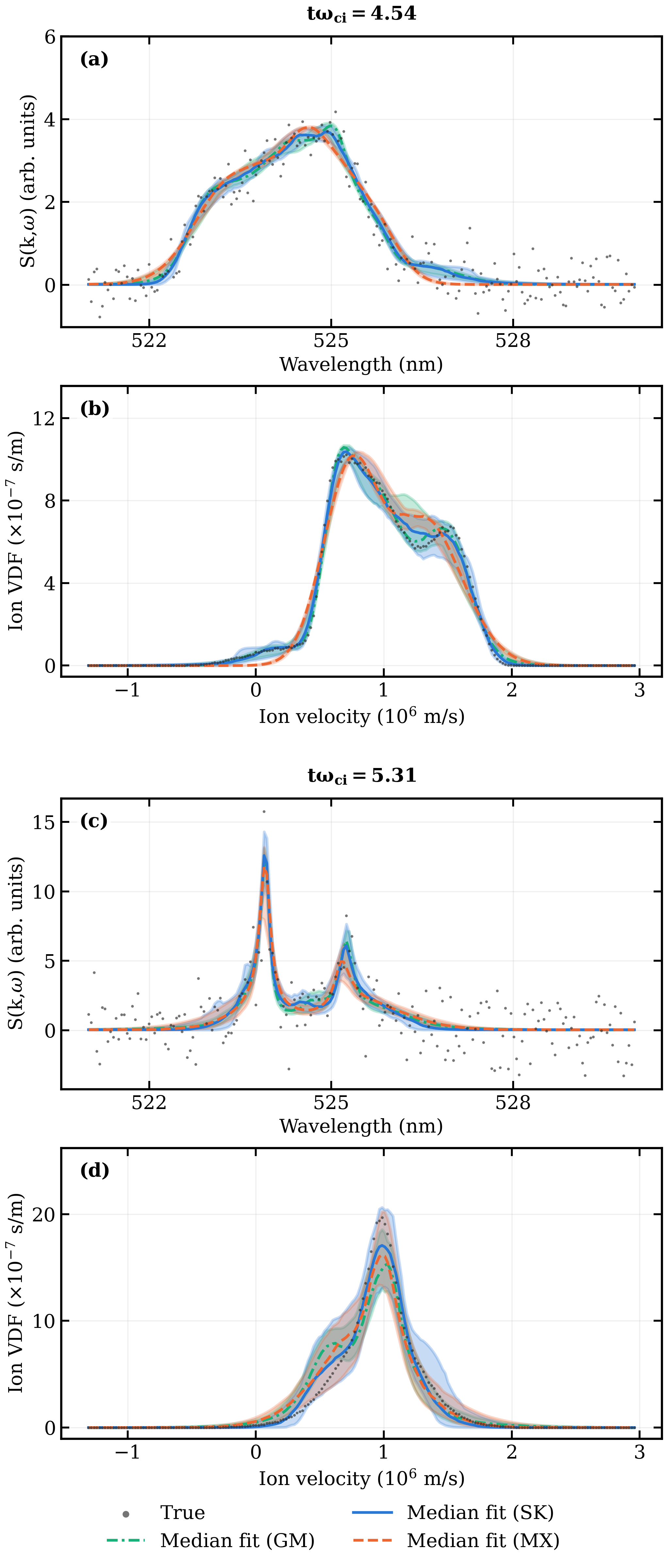}
    \caption{
        Monte Carlo results for two times. (a)/(b) True IAW spectrum/VDF for $t\omega_{ci}=4.54$ (points) with median/uncertainty (line/shaded region) for each basis overlaid; (c)/(d) True IAW spectrum/VDF $t\omega_{ci}=5.31$ with median/uncertainty for each basis overlaid. Median and uncertainty (5th/95th percentile values) are computed pointwise within the primary group.
    }
    \label{fig:mc_spectra_vdf_grid}
\end{figure}

\begin{figure}[!htbp]
    \centering
    \includegraphics[width=\columnwidth]{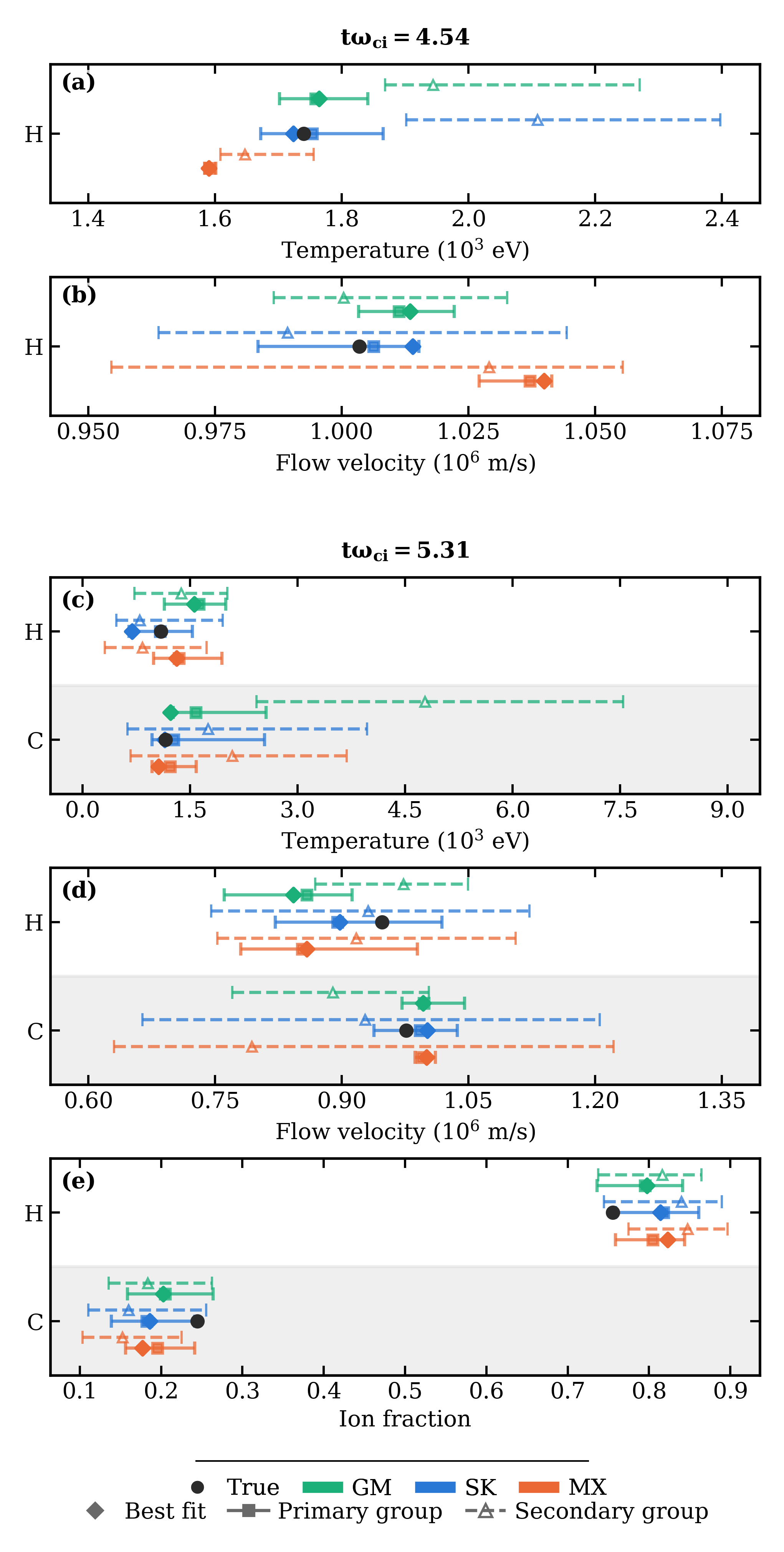}
    \caption{
        Per-group moments with uncertainties for two times. (a) Hydrogen temperature; (b) Hydrogen flow velocity; (c) H/C temperature; (d) H/C flow velocity; (e) H/C relative weights. Original best fit points are plotted as diamonds, primary group 5th/50th/95th percentile values are plotted solid, and secondary group 5th/50th/95th percentile values are plotted dashed. Colors correspond to basis.
    }
    \label{fig:mc_moment_comparison}
\end{figure}

With clearly identified groups, the per-group statistics are reported as the 5th, 50th, and 95th percentile values for all quantities. Note that these percentiles describe the spread of converged solutions across the perturbed initializations, all fit to the same noise realization. As such, the ``uncertainty'' reported here is a statement on initialization dependence and does not include noise-based uncertainty. The primary group spectra and VDFs are plotted in Figure~\ref{fig:mc_spectra_vdf_grid}, and the computed moments for both groups are plotted in Figure~\ref{fig:mc_moment_comparison}.

At $t\omega_{ci}=4.54$, the true VDF is captured within uncertainty for GM and SK, while MX fails to do the same, highlighting the inability of a Maxwellian model to reconstruct this non-thermal shocked plasma. Additionally, the median fits of SK and especially GM track more closely to the ground truth VDF than the original best fit (see Figure~\ref{fig:1d_ion_vdf_grid}), suggesting that it is possible for the median of a large sample of random initializations to reconstruct the VDF better than any individual fit.

The fit at $t\omega_{ci}=5.31$ behaves differently: all methods perform similarly, with a notable leftward shift in the VDF base compared to the ground truth. The common shift across all methods points to a spectral inversion artifact rather than a basis-specific defect, further supported by the very similar spectral reconstruction across all methods. Furthermore, SK and MX both capture the true VDF within uncertainties, while GM misses it entirely on both the left side and the peak. We suspect this is due to independently perturbing each component of the multi-component population.

Figure~\ref{fig:mc_moment_comparison} depicts the 5th/50th/95th percentile moments of both primary and secondary groups for each basis method, with the initial best fit and ground truth values for reference. At $t\omega_{ci}=4.54$, both the GM and SK primary groups capture the true temperature and flow velocity within uncertainty, with the median in both cases being a better reconstruction than the initial best fit. The MX primary group is more tightly clustered in both temperature and flow velocity than GM or SK, yet consistently under- and overestimates the respective moments. This allows us to more confidently conclude that MX is a fundamentally less-suited model for this time, as expected for a non-thermal ground truth VDF.

The secondary groups for $t\omega_{ci}=4.54$ have a consistently larger spread across all bases and moments, with the medians noticeably different than those of the primary group. Interestingly, the secondary group for MX captures the true temperature within its uncertainty. However, this does not suggest that MX is a good model here: the low uncertainty of the primary group points to a clear minimum in the loss function, and the median loss for the secondary group is 10\% larger, indicating a group largely composed of less-converged fits. The rightward skew of the 95th percentile further suggests that this is dominated by hotter outliers; a closer look at these specific hot VDFs showed, in many cases, a single-peak structure that is not representative of the true VDF. So MX can occasionally capture the true temperature, but through temperature's definition as an integrated moment of the VDF rather than through accurate phase-space reconstruction---unlike GM and SK.

At $t\omega_{ci}=5.31$, the primary group for SK displays the strongest overall recovery of the ground truth moments, with all values being within uncertainties. Similarly, the primary MX group captures all moments except for the carbon flow velocity, which is tightly constrained around $9.95\times10^5$ m/s, within $\sim$2\% of the true value. Notably, GM only captures the carbon flow velocity within uncertainty; all other GM moments fall slightly outside of the 5th/95th percentile window.

All methods at $t\omega_{ci}=5.31$ display the same moment trends across primary and secondary groups to varying degrees: the secondary group median shows hotter and slower carbon, colder and faster hydrogen, and a lower carbon fraction than the primary group median. This better characterizes the mechanism of the H/C degeneracy---the optimizer tends to find both a slower and faster population (evident in the primary flow velocity groups in Figure~\ref{fig:mc_moment_comparison}~(d)) and is effectively swapping between hydrogen and carbon for each group. The temperature changes can be explained by the larger mass of carbon compared to hydrogen, which causes the VDF of a hotter carbon population to resemble that of a cold hydrogen population. The two can therefore be interchanged with effectively no spectral consequence, provided the temperatures adjust accordingly---which is what we observe at this time step.

Furthermore, the median NMSE for the $t\omega_{ci}=5.31$ primary groups is within a few percent of the initial best fit. The secondary groups, however, reveal additional structure about the loss landscape for each method. For SK and MX, the secondary group has a median loss 20\% and 21\% greater than their primary groups, respectively, whereas this difference is near-zero for GM. Taking this and the moment trends into account provides a more nuanced view of the H/C degeneracy---it exists regardless of basis as a property of the inversion problem for a multi-ion plasma, yet the degenerate groups are not equally well-fit across bases. For SK and MX, the degenerate secondary group is a noticeably worse-fit set of solutions, whereas for GM both groups represent a truly degenerate solution set where NMSE does not serve to distinguish the two solutions.

 We suspect the discrepancies with GM are due to an independent perturbation to the less-constrained components of the multi-component VDF. The MC applies a perturbation to the mean of each component, injecting internal spread. This would compound with the inherently less-constrained inversion of multi-ion plasmas to inflate the temperature, which is observed in Figure~\ref{fig:mc_moment_comparison}---the initial best fit carbon temperature is below the 5th percentile of the primary group, the primary group's median is noticeably larger than that of SK or MX, and the secondary group's carbon temperature is drastically inflated. This effect is primarily seen in the carbon population due to its lower relative weight, as even significant changes to a lower-weighted population can have minimal effects on the calculated TS spectrum. A more targeted MC that does not perturb each component independently would likely reduce this temperature inflation.

This method of uncertainty quantification reinforces the earlier observations of improved fit quality in non-thermal regions and gives a confidence level for each fit, while revealing structure in the inversion and the loss landscape that no single fit can expose. A large statistical sample is thus an effective way to assess fit quality and uncertainty, and it is possible for its median to reconstruct the VDF better than any individual fit.

\section{Conclusions and Future Work}
\label{sec:conclusion}

In this work, we developed a generalized Thomson scattering analysis framework for recovering complex VDFs from collective Thomson scattering spectra. Representing electron and ion distributions with either a superposition of Gaussian components or a generalized Super-Kappa distribution, and pairing these parameterizations with a gradient-descent optimization routine, yields a flexible inversion that recovers non-thermal distributions while retaining comparable performance in simpler regions. We evaluated the approach using synthetic TS spectra generated from PIC simulations of collisionless shock systems containing interacting hydrogen and carbon ion populations.

Both basis representations reproduced the dominant EPW and IAW spectral features throughout the analysis interval and recovered the major characteristics of the underlying electron and ion VDFs, demonstrating notable improvement over the simpler Maxwellian model in regions with highly non-thermal VDFs. The reconstructed plasma moments agreed well with the simulation values: electron moments were recovered to within 2--5\% for all methods and times, while ion moments carried larger errors---2--5\% on flow velocity and 10--17\% on temperature---and larger differences between bases.

We have also outlined a Monte Carlo-based method of uncertainty quantification that provides further insight into the multi-modal loss landscape for complex fits. For the two times that were analyzed, this method revealed two solution groups, attributed to multi-ion degeneracies and the inherent complexity of the inversion problem. Splitting by group gave a more accurate uncertainty on the extracted VDFs and moments; applying it at additional times and to spectra of varying complexity would further map the solution landscape for TS inversion.

We do not intend for these methods to be the definitive solution for TS analysis; rather, their performance demonstrates the feasibility of using various generalized basis representations to improve fit quality in regions where non-thermal plasmas are expected. There is significant room for exploration in this space to find robust ways of modeling the underlying VDFs. Furthermore, the Monte Carlo method and GMM-based grouping can provide higher confidence in extracted parameters and serve as an intuitive uncertainty metric by construction.

We suspect that future TS analysis will involve adjusting the complexity of the basis representation according to the observed and/or predicted plasma complexity at a given time or location. In this way, regions that require non-thermal bases would get the flexibility needed to accurately reconstruct the VDFs, whereas simpler fits would not suffer from unnecessary internal degeneracies or overfitting. A BIC-style selection mechanism~\cite{schwarz1978}, analogous to the GMM $k$-selection in Section~\ref{sec:uncertainty}, could choose that complexity quantitatively for each fit. Angularly resolved TS~\cite{katz2024} offers an independent handle on VDF shape and could offer a route to breaking the H/C degeneracy identified here.

\begin{acknowledgments}
This work was supported by the DOE NNSA under Award No. DE-NA0004033, by the DOE FES under Award No. DE-SC0024566, by the NNSA Center of Excellence under cooperative agreement number DE-NA0004146, and by NASA under Grant No. 80NSSC23K1035. This work was also supported by the UCLA Physics and Astronomy Department REU Program. This research used resources of the Oak Ridge Leadership Computing Facility, which is a DOE Office of Science User Facility supported under Contract DE-AC05-00OR22725.\end{acknowledgments}

\section*{Author Declarations}
\subsection*{Conflict of Interest}
The authors have no conflicts to disclose.

\subsection*{Author Contributions}
\textbf{Greg A. Baker:} Conceptualization (supporting); Formal analysis (equal);
Investigation (equal); Methodology (equal); Software (equal);
Validation (equal); Visualization (equal); Writing -- original draft
(supporting); Writing -- review \& editing (lead).
\textbf{Bradyn Klein:} Conceptualization (equal); Data curation (supporting);
Formal analysis (equal); Investigation (equal); Methodology (equal);
Software (equal); Validation (equal); Visualization (equal);
Writing -- original draft (lead); Writing -- review \& editing (supporting).
\textbf{Kristopher Wright:} Conceptualization (supporting); Formal analysis
(supporting); Methodology (supporting); Writing -- review \& editing
(supporting).
\textbf{Vedang Bhelande:} Methodology (supporting); Software (equal);
Writing -- review \& editing (supporting).
\textbf{Zhaoyan Sun:} Data curation (supporting); Investigation (supporting);
Writing -- review \& editing (supporting).
\textbf{Rayner Sutardja:} Data curation (supporting); Writing -- review \&
editing (supporting).
\textbf{Samuel R. Totorica:} Data curation (lead); Writing -- review \& editing (supporting).
\textbf{Derek B. Schaeffer:} Conceptualization (equal); Funding acquisition
(lead); Project administration (lead); Resources (lead); Supervision (lead);
Writing -- review \& editing (supporting).

\section*{Data Availability}
The data that support the findings of this study are available from
the corresponding author upon reasonable request.

\bibliographystyle{apsrev4-2}
\bibliography{refs.bib}

\end{document}